\documentclass[prd,twocolumn,floatfix,amsmath,nofootinbib,amssymb,floatfix]{revtex4}
\usepackage{graphicx,color,dcolumn,booktabs,bm}
\usepackage{longtable,lscape}
\usepackage{pdfpages}
\usepackage{txfonts}
\usepackage{overpic}
\usepackage{amssymb}
\usepackage{makecell}
\usepackage{indentfirst}
\usepackage{feynmf}   
\usepackage{slashed}  
\usepackage{cases}
\usepackage{color}
\usepackage{multirow}
\usepackage{threeparttable}
\usepackage{epstopdf}
\usepackage{enumerate}
\usepackage{subfigure}
\usepackage{diagbox}
\usepackage{graphicx,color,dcolumn,booktabs,bm}
\usepackage{mathrsfs}
\usepackage{cancel}
\usepackage{float}
\usepackage[colorlinks,
            citecolor=blue,
            anchorcolor=red,
            menucolor=red,
            linkcolor=red,
            filecolor=red,
            runcolor=red,
            urlcolor=blue,
            frenchlinks=red]{hyperref}

\begin{document}
\title{Mass spectra of doubly heavy baryons in the relativized quark model with heavy-quark dominance}
\author{Zhen-Yu Li$^{1,4}$}
\email{zhenyvli@163.com }
\author{Guo-Liang Yu$^{2}$}
\email{yuguoliang2011@163.com }
\author{Zhi-Gang Wang$^{2}$ }
\email{zgwang@aliyun.com }
\author{Jian-Zhong Gu$^{3}$ }
\email{gujianzhong2000@aliyun.com }
\author{Hong-Tao Shen$^{4}$ }
\email{shenht@gxnu.edu.cn }

\affiliation{$^1$ School of Physics and Electronic Science, Guizhou Education University, Guiyang 550018,
China\\$^2$ Department of Mathematics and Physics, North China Electric Power University, Baoding 071003,
China\\$^3$ China Institute of Atomic Energy, Beijing 102413,China\\$^4$ Guangxi Key Lab of Nuclear Physics and Technology, Guangxi Normal University, Guilin 541006, China}
\date{\today }

\begin{abstract}
In the framework of the relativized quark model, the mass spectra of the doubly heavy baryons are rigorously calculated in the three-quark system under the heavy-quark dominance mechanism, by using the Gaussian expansion method and the infinitesimally-shifted Gaussian basis functions. With the obtained mass spectra of all doubly heavy baryon families, the contribution of each Hamiltonian term to the energy levels is analyzed. It is found that the spin splitting is mainly determined by the spin-dependent interactions associated with the light quark. Moreover, it is shown that the spin splitting evolves regularly with the mass of heavy quarks by the evolution of the spectral structure, which is consistent with the heavy quark symmetry. Meanwhile, the orbital excitation is dominated by the $\rho$-mode, which is different from that of the singly heavy baryons. At last, our analysis indicates that the $\Xi_{cc}^{+}(3520)$ state should not exist truly and the $\Xi_{cc}^{++}(3621)$ should be the true ground state with $J^{P}$ = $\frac{1}{2}^{+}$. It is recommended to design the corresponding experiments to search for the $\Xi_{cc}^{*}$ in the energy range from 3694 to 3714 MeV.

\end{abstract}

\maketitle

\section{Introduction}\label{sec1}

The heavy baryon spectroscopy has always been a hot topic in the study of the intra-hadron dynamics of non-perturbative QCD and confinement~\cite{01,01p,02,03}. In recent decades, a large number of singly heavy baryons have been experimentally observed~\cite{04}, which has strongly promoted the related theoretical research~\cite{05,06}.
While, experimentally searching for the doubly heavy baryons (DHBs) presents a different scene.
In 2002, a signal of the $\Xi_{cc}^{+}(3520)$ was first reported in the $\Lambda_{c}^{+}K^{-}\pi^{+}$ final state by the SELEX collaboration~\cite{07}. But the confirmation of this state has not yet reached a conclusion, and searching for the DHBs came to a standstill~\cite{08,09,10,11,12}. The breakthrough came in 2017 with the discovery of a doubly charmed baryon $\Xi_{cc}^{++}(3621)$ by the LHCb collaboration, in the decay mode $\Xi_{cc}^{++}\rightarrow \Lambda_{c}^{+}K^{-}\pi^{+}\pi^{+}$~\cite{13,14}. Later, the LHCb collaboration reconfirmed this baryon in another decay mode $\Xi_{cc}^{++}\rightarrow \Xi_{c}^{+}\pi^{+}$~\cite{15}. The efforts of searching for the $\Xi_{bc}^{0}$~\cite{16}, $\Omega_{bc}^{0}$~\cite{17} and $\Xi_{bc}^{+}$~\cite{18} baryons were reported one after another later on. But, no evidence of the signals was found. Up to now, only the $\Xi_{cc}^{++}(3621)$ and $\Xi_{cc}^{+}(3520)$ have been collected in the new Review of Particle Physics (RPP) by the Particle Data Group (PDG).

For promoting the experimental research of the DHBs, the theoretical studies have provided many valuable clues, such as the predicted production~\cite{19,20,21,22,23,24,25,26}, the decay feature~\cite{27,28,29,30,31,32,33,33p,34,34p,35}, the magnetic moments~\cite{36,37,38,39,40}, as well as the spectral structures from various methods including the lattice QCD~\cite{41,42,43,44,45,46,46pp,46p,46p2}, the QCD sum rules~\cite{47,48,49,50,51,52,53,54}, the Bethe-Salpeter equation~\cite{55,56,57,58}, all kinds of the quark potential models~\cite{62,62p1,62p2,62p3,62p4,62p5,62p6,62p7,62p8,62p9,62p10,62p11,62p12,62p13,62p14,62p15,62p16,62p17,62p18} and other methods~\cite{63,64}.

Nevertheless, the accuracy of theoretical calculation has always been a difficult problem. Some predicted mass values of the $\Xi_{cc}$ and $\Xi_{cc}^{*}$ baryons in different years over the past two decades are listed in Table~\ref{a01}. It can be found that the theoretical values fluctuate significantly over the years. As is shown in Table~\ref{a01}, the predicted values $M_{\Xi_{cc}}$ were mainly around 3.62 GeV before 2002. From 2004 to 2014, however, they dropped to around 3.55 GeV, which was likely influenced by the 2002 report from the SELEX collaboration on the $\Xi_{cc}^{+}(3520)$. After the $\Xi_{cc}^{++}(3621)$ was discovered with a high statistical significance, the predicted mass values $M_{\Xi_{cc}}$ returned to around 3.62 GeV. It is worth noting that from 2014 to 2017 some theoretical methods independently obtained relatively good calculation results, especially the lattice QCD~\cite{44}.  In contrast, the performance of various quark potential models was less than ideal, except for the work of Ebert et al. in 2002. So, it is necessary to investigate whether the quark potential models can provide accurate predictions in the study of the DHBs.
\begin{table*}[htbp]
\begin{ruledtabular}\caption{Predicted mass central values (in GeV) of the ground states of the double charm baryons in some references. The superscript * refers to the $J$ = 3/2 baryons. }
\label{a01}
\begin{tabular}{c c c c | c c c c  }
$M_{\Xi_{cc}}$ & $M_{\Xi_{cc}^{*}}$ & Method & Year  & $M_{\Xi_{cc}}$ & $M_{\Xi_{cc}^{*}}$ & Method & Year   \\ \hline
3.63 & -  & Potential model & 1994~\cite{62}  & 3.627 & 3.690  & Empirical Formula & 2014~\cite{62p9}   \\
3.61 & 3.68 & One-Gluon-Exchange Model & 1994~\cite{01p}  & 3.615 & 3.747  & Diquark Model & 2017~\cite{62p10}  \\
3.660 & 3.74  & Feynman-Hellman Theorem & 1995~\cite{63} & 3.606 & 3.675  & Diquark Model & 2017~\cite{30}   \\
3.66 & 3.81  & Diquark Model & 1996~\cite{62p1}  & 3.66 & -  & QCD Sum Rules & 2017~\cite{46pp}     \\
3.478 & 3.61  & Diquark Model & 2000~\cite{62p2}  & 3.627 & 3.690  & Diquark Model & 2018~\cite{62p11}    \\
3.620 & 3.727  & Diquark Model & 2002~\cite{62p3}  & 3.626 & 3.693  & Lattice QCD & 2018~\cite{46p}    \\
3.55 & 3.59  & Bag model & 2004~\cite{62p4}  &3.522 & 3.696  & Hypercentral Potential & 2018~\cite{62p12}   \\
3.51 & 3.548  & Quark Model & 2007~\cite{62p5}  & 3.63 & 3.75  & QCD Sum Rules & 2018~\cite{54}   \\
3.519 & 3.555  & Diquark Model & 2008~\cite{62p6}  & 3.54 & 3.62  & Bethe-Salpeter Equation & 2018~\cite{58}   \\
3.547 & 3.719  & Salpeter Model & 2009~\cite{56}  & 3.601 & 3.703  & Diquark Model & 2019~\cite{62p13}    \\
3.678 & 3.752  & Diquark Model & 2012~\cite{62p7}  & 3.654 & 3.741  & Effective String  & 2021~\cite{64}    \\
3.532 & 3.623  & Non-relativistic Quark Model & 2014~\cite{62p8}  & 3.604 & 3.714  & Bag Model  & 2021~\cite{62p14}    \\
3.610 & 3.692  & Lattice QCD & 2014~\cite{44}  & 3.640 & 3.695  & Relativized Quark Model  & 2022~\cite{62p17}   \\
\end{tabular}
\end{ruledtabular}
\end{table*}

As a representative quark potential model, the relativized quark model (RQM) has gone through a tortuous development process and now achieved further success in the studies of the singly heavy baryon spectroscopy.
The RQM was first applied to study the meson spectroscopy by Godfrey and Isgur in 1985~\cite{F401}. The Hamiltonian of this model is based on a confining potential and an effective one-gluon exchange motivated by QCD and contains almost all possible forms of the main interactions between two quarks and has a unique advantage in describing the strong interactions within the hadron in detail. In 1986, Capstick and Isgur extended this model and tried to describe the mass spectra of all baryons within a unified framework~\cite{F402}. Their work has been a guideline for experimenters and theorists~\cite{01,Fp003}. Nevertheless, they only demonstrated a part of the great strength of this model.
The literature suggests that the outstanding advantages of this model have not been fully appreciated in the nearly 30 years since it was proposed~\cite{02,03}. This lies in the two main reasons summarized here. First, the interactions within a heavy baryon have not been well understood yet. Second, the rigorous calculation of this model for a baryon system was difficult and some approximations had to be adopted, such as the quark-diquark approximation~\cite{88}.

Actually, the concept of a diquark has always been useful for understanding the hadron structure and the high-energy particle reactions~\cite{89}, even though it has not been fully accepted yet~\cite{01}. The quark-diquark picture was applied to the heavy baryons very early. Later in 2011~\cite{F405}, Ebert, Faustov and Galkin successfully solved the problem of the missing states~\cite{02} in the singly heavy baryons by using the heavy quark-light diquark picture, and reproduced almost all the observed singly heavy baryons at that time. In their work,  all heavy baryon excitations, both orbital and radial excitations, were assumed to only occur in the bound system of the heavy quark and light diquark, which is very similar to the orbital excitation of the $\lambda$-mode in the genuine three-quark picture of a baryon. Inspired by this similarity and some pioneering works in the past~\cite{02,91,Isgur1991,F403}, it was gradually recognized that within a singly heavy baryon, the orbital excitation in the three-quark picture is dominated by the $\lambda$-mode~\cite{91,94,95}, but the $\rho$-mode within a doubly heavy baryon~\cite{62p16,62p17,62p18}. Later, it was understood as the mechanism of the heavy-quark dominance (HQD), which may determine the overall structure of the excitation spectra for the singly and doubly heavy baryons and solve the problem of the `missing' states in a natural way~\cite{96}.

On the other hand, the Gaussian expansion method (GEM) and the infinitesimally-shifted Gaussian (ISG) basis functions~\cite{F6021,F602} have been introduced into the multi-quark system these years~\cite{99,100}. These techniques have taken a substantial step forward in the rigorous calculation of the baryon systems.

Based on the above progresses, very recently, the mass spectra of the singly heavy baryons have been obtained rigorously in the genuine three-quark picture, in the framework of the RQM and the HQD mechanism, by using the GEM and the ISG basis functions~\cite{101}.
It turns out that the combination method of the RQM + HQD + GEM(ISG) used in the singly heavy baryons is successful and reliable in reproducing the data and effectively clarifying some fundamental issues. Therefore, extending this method to the study of the DHBs will further test the application scope of this combination method and also answer the questions mentioned above, i.e., whether and how the RQM can provide accurate predictions in the DHBs spectroscopy. Additionally, it can provide a theoretical support for the current experimental research of the DHBs.

The remainder of this paper is organized as follows. In
Sec.~\ref{sec2}, the theoretical methods used in this work are introduced briefly, including the Hamiltonian of the relativized quark model, the wave functions and the Jacobi coordinates, as well as the GEM and the ISG basis functions. The structural properties of the DHBs spectra, the reliability of the combination method, and the search for the DHBs are analyzed in Sec.~\ref{sec3}. And Sec.~\ref{sec4} is reserved for the conclusions.

\section{Theoretical methods used in this work}\label{sec2}
This method has been described in detail in the study of the singly heavy baryons~\cite{101}. Here is its brief introduction for the convenience of discussion.
In the relativized quark model, the Hamiltonian for a three-quark system is based on the two-body interactions,
 \begin{eqnarray}
 \begin{aligned}
&H=H_{0}+\sum_{i<j}V_{ij},\\
&H_{0}=\sum_{i=1}^{3}\sqrt{p_{i}^{2}+m_{i}^{2}},\\
&V_{ij}=\tilde{H}^{conf}_{ij}+\tilde{H}^{hyp}_{ij}+\tilde{H}^{so}_{ij},
\end{aligned}
\end{eqnarray}
where the position-related term $\tilde{H}^{conf}_{ij}$ is the confinement interaction. The spin-dependent terms $\tilde{H}^{hyp}_{ij}$ and $\tilde{H}^{so}_{ij}$ are the hyperfine and spin-orbit interactions, respectively.
The confinement term $\tilde{H}^{conf}_{ij}$ includes a modified one-gluon-exchange potential $G'_{ij}(r)$ and a smeared linear confinement potential $\tilde{S}_{ij}(r)$. The hyperfine interaction $\tilde{H}^{hyp}_{ij}$ consists of the tensor term $\tilde{H}^{t}_{ij}$ and the contact term $\tilde{H}^{c}_{ij}$. And the spin-orbit interaction $\tilde{H}^{so}_{ij}$ can be divided into the color-magnetic term $\tilde{H}^{so(v)}_{ij}$ and the Thomas-precession term $\tilde{H}^{so(s)}_{ij}$.
They are expressed as
 \begin{eqnarray}
 \begin{aligned}
&\tilde{H}^{conf}_{ij}=G'_{ij}(r)+\tilde{S}_{ij}(r), \\
&\tilde{H}^{hyp}_{ij}=\tilde{H}^{t}_{ij}+\tilde{H}^{c}_{ij},\\
&\tilde{H}^{so}_{ij}=\tilde{H}^{so(v)}_{ij}+\tilde{H}^{so(s)}_{ij},
\end{aligned}
\end{eqnarray}
with
\begin{eqnarray}
 &\tilde{H}^{t}_{ij}=-\frac{\textbf{s}_{i}\cdot\textbf{r}_{ij}\textbf{s}_{j}\cdot\textbf{r}_{ij}/r^{2}_{ij}-\frac{1}{3}\textbf{s}_{i}\cdot\textbf{s}_{j}}{m_{i}m_{j}}
\times(\frac{\partial^{2}}{\partial{r^{2}_{ij}}}-\frac{1}{r_{ij}}\frac{\partial}{\partial{r_{ij}}})\tilde{G}^{t}_{ij}, \\
&\tilde{H}^{c}_{ij}=\frac{2\textbf{s}_{i}\cdot\textbf{s}_{j}}{3m_{i}m_{j}}\nabla^{2}\tilde{G}^{c}_{ij},\\
\label{e5}
&\tilde{H}^{so(v)}_{ij}=\frac{\textbf{s}_{i}\cdot\textbf{L}_{(ij)i}}{2m^{2}_{i}r_{ij}}\frac{\partial\tilde{G}^{so(v)}_{ii}}{\partial{r_{ij}}}+
\frac{\textbf{s}_{j}\cdot\textbf{L}_{(ij)j}}{2m^{2}_{j}r_{ij}}\frac{\partial\tilde{G}^{so(v)}_{jj}}{\partial{r_{ij}}}+
\frac{(\textbf{s}_{i}\cdot\textbf{L}_{(ij)j}+\textbf{s}_{j}\cdot\textbf{L}_{(ij)i})}{m_{i}m_{j}r_{ij}}\frac{\partial\tilde{G}^{so(v)}_{ij}}{\partial{r_{ij}}}, \\
\label{e6}
&\tilde{H}^{so(s)}_{ij}=-\frac{\textbf{s}_{i}\cdot\textbf{L}_{(ij)i}}{2m^{2}_{i}r_{ij}}\frac{\partial\tilde{S}^{so(s)}_{ii}}{\partial{r_{ij}}}-
\frac{\textbf{s}_{j}\cdot\textbf{L}_{(ij)j}}{2m^{2}_{j}r_{ij}}\frac{\partial\tilde{S}^{so(s)}_{jj}}{\partial{r_{ij}}}.
\end{eqnarray}
Here, the following conventions are used, $\textbf{L}_{(ij)i}=\mathbf{r}_{ij}\times\mathbf{p}_{i}$ and $\textbf{L}_{(ij)j}=-\mathbf{r}_{ij}\times\mathbf{p}_{j}$.
All of the interaction terms have been modified with the momentum-dependent factors. For a baryon state, the color interaction between any two quarks can be simply expressed as a -2/3 factor. The detailed description of this model can be found in Refs.~\cite{F401,F402}.
The used parameters in the model are the same as those in our previous work~\cite{101}.

The full wave function of a doubly heavy baryon can be expressed as the direct product of the color wave function, flavor wave function and spatial wave function. Here, the two heavy flavor quarks are assumed to form a subsystem with their exchange symmetry. Then, the flavor wave functions are written as
\begin{eqnarray}
\begin{aligned}
&\Xi_{cc}^{++}=(cc)u,~\Xi_{cc}^{+}=(cc)d, \\
&\Xi_{bc}^{+}=\frac{1}{\sqrt{2}}(bc+cb)u,~\Xi_{bc}^{0}=\frac{1}{\sqrt{2}}(bc+cb)d,\\
&\Xi_{bb}^{0}=(bb)u,~\Xi_{bb}^{-}=(bb)d,\\
&\Xi_{bc}^{'+}=\frac{1}{\sqrt{2}}(bc-cb)u,~\Xi_{bc}^{'0}=\frac{1}{\sqrt{2}}(bc-cb)d,
\end{aligned}
\end{eqnarray}
for the non-strange DHBs, and
\begin{eqnarray}
\begin{aligned}
&\Omega_{cc}^{+}=(cc)s, \\
&\Omega_{bc}^{0}=\frac{1}{\sqrt{2}}(bc+cb)s,\\
&\Omega_{bb}^{-}=(bb)s,\\
&\Omega_{bc}^{'0}=\frac{1}{\sqrt{2}}(bc-cb)s,
\end{aligned}
\end{eqnarray}
for the strange ones. Here $u$, $d$, $s$, $b$ and $c$ denote up, down, strange, bottom and charm quarks, respectively.

\begin{figure}[htbp]
\centering
\includegraphics[width=8.5cm]{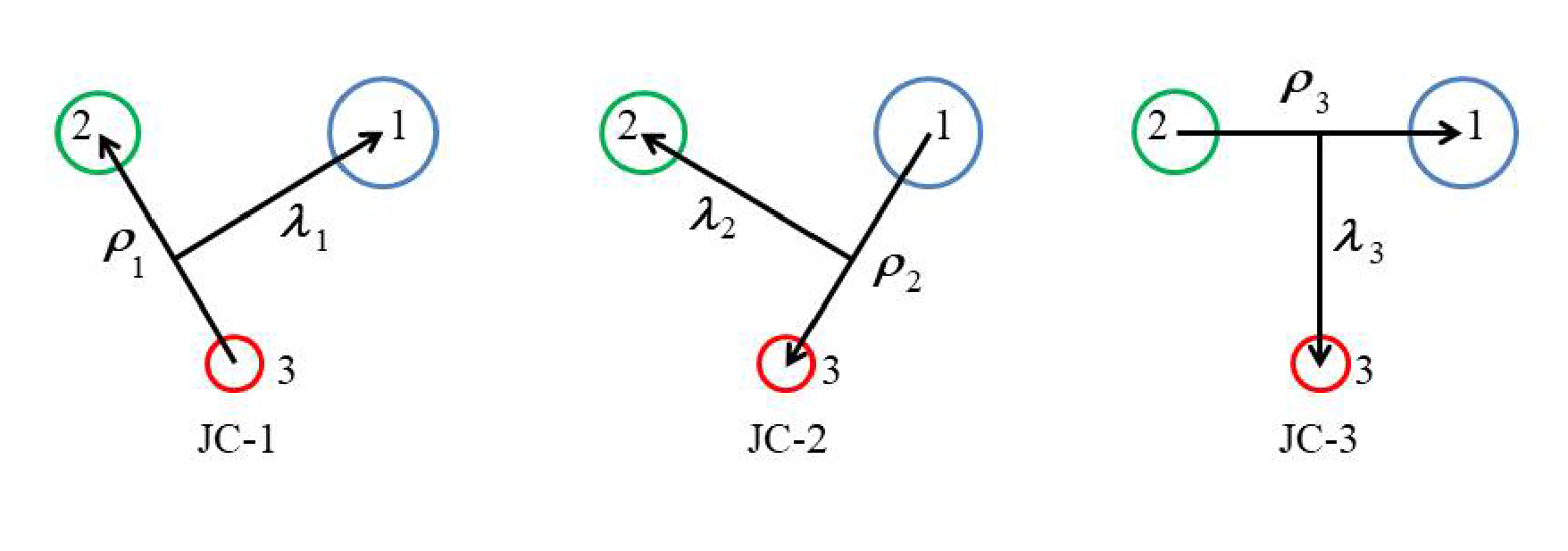}
\caption{There are 3 channels of the Jacobi coordinates for a three-quark system. The channel 3 (JC-3) is selected for defining the spatial wave function of a doubly heavy baryon state. All the quarks are numbered for ease of use in calculations, and the 3rd quark refers specifically to the light quark.}
\label{f1}
\end{figure}

For describing the internal orbital motion, the specific Jacobi coordinates (named JC-3 for short) can be selected as shown in Fig.~\ref{f1}, which is consistent with the above reservation about the flavor wave function naturally.
The Jacobi coordinates are defined as
\begin{eqnarray}
\begin{aligned}
&\boldsymbol\rho_{i}=\textbf{r}_{jk}=\textbf{r}_{j}-\textbf{r}_{k}, \\
&\boldsymbol\lambda_{i}=\textbf{r}_{i}-\frac{m_{j}\textbf{r}_{j}+m_{k}\textbf{r}_{k}}{m_{j}+m_{k}},\\
&\boldsymbol R_{i}=\frac{m_{i}\textbf{r}_{i}+m_{j}\textbf{r}_{j}+m_{k}\textbf{r}_{k}}{m_{i}+m_{j}+m_{k}}\equiv \mathbf{0},
\label{e13}
\end{aligned}
\end{eqnarray}
where $\{i$, $j$, $k\}$ = $\{$1, 2, 3$\}$, $\{$2, 3, 1$\}$ or $\{$3, 1, 2$\}$. $\textbf{r}_{i}$ and $m_{i}$ denote the position vector and the mass of the $i$th quark, respectively. $\textbf{\emph{R}}_{i}\equiv \textbf{0}$ means that the kinetic energy of the center of mass is not considered. Specially, for the JC-3 in Fig.~\ref{f1}, the following definitions are used, $\boldsymbol\rho_{3}\equiv \boldsymbol\rho$ and $\boldsymbol\lambda_{3}\equiv \boldsymbol\lambda$.

In the JC-3, the spatial (spin and orbit) wave function is assumed to have the coupling scheme
\begin{eqnarray}
|(J^{P})_{j},L\rangle = |\{[(l_{\rho} l_{\lambda} )_{L}(s_{1}s_{2})_{s_{12}}]_{j} s_{3}\}_{J }\rangle,
\end{eqnarray}
with the Parity number $P=(-1)^{l_{\rho}+l_{\lambda}}$.
$l_{\rho}$($l_{\lambda}$), $L$ and $s_{12}$ are the quantum numbers of the relative orbital angular momentum $\textbf{\emph{l}}_{\rho}$ ($\textbf{\emph{l}}_{\lambda}$), total orbital angular momentum $\textbf{\emph{L}}$, and total spin of the light-quark pair $\mathbf{s}_{12}$, respectively. $j$ denotes the quantum number of the coupled angular momentum of $\textbf{\emph{L}}$ and $\textbf{s}_{12}$, so that the total angular momentum $J=j\pm\frac{1}{2}$.
More precisely, the baryon state is labeled with $(l_{\rho},l_{\lambda})nL(J^{P})_{j}$, in which $n$ is the quantum number of the radial excitation.
For the $\Xi_{bc}^{'}$ and $\Omega_{bc}^{'}$ families, $(-1)^{l_{\rho}+s_{12}}=1$ should be also guaranteed due to the total antisymmetry of the two heavy quarks, but  $(-1)^{l_{\rho}+s_{12}}=-1$ for the other DHBs. All the conventions are based on the JC-3 in Fig.~\ref{f1}.

The numerical calculations are carried out based on the GEM and the ISG basis functions.
Given a set of the orbital quantum numbers $\{l$, $m\}$, the Gaussian basis function $|(nlm)^{G} \rangle$ is commonly written in  position space as
 \begin{eqnarray}
\begin{aligned}
 &\phi^{G}_{nlm}(\textbf{r})=\phi^{G}_{nl}(r)Y_{lm}(\hat{\textbf{r}}),\\
 &\phi^{G}_{nl}(r)=N_{nl}r^{l}e^{-\nu_{n}r^{2}},\\
 &N_{nl}=\sqrt{\frac{2^{l+2}(2\nu_{n})^{l+3/2}}{\sqrt{\pi}(2l+1)!!}},
\end{aligned}
\end{eqnarray}
with
\begin{eqnarray}
\begin{aligned}
& \nu_{n}=\frac{1}{r^{2}_{n}},\\
& r_{n}=r_{1}a^{n-1}\ \ \ (n=1,\ 2,\ ...,\ n_{max}).
\end{aligned}
\end{eqnarray}
$\{r_{1}, a, n_{max}\}$ (or equivalently $\{n_{max},r_{1},r_{n_{max}}\}$) are the Gaussian size
parameters and commonly related to the scale in question~\cite{F602}. In general, if the $r_{1}$ and $r_{n_{max}}$ are properly valued, the convergence of the GEM can be very rapid and the required precision can be guaranteed~\cite{94,F602}. Through continuous attempts, the $r_{1}$ and $r_{n_{max}}$ are eventually taken as 0.18 GeV$^{-1}$ and 15 GeV$^{-1}$, respectively. In this case, the expected value of the Hamiltonian converges rapidly as the dimension number $n_{max}$ increases, and remains stable when $n_{max}$ is in the range of 9 to 14. Finally, the optimized values of $\{n_{max}=10$, $r_{1}=0.18$ GeV$^{-1}$, $r_{n_{max}}=15$ GeV$^{-1}\}$ are selected for the DHBs in this work. Details can be found in Refs.~\cite{F502,F503}.

For the doubly heavy baryon system, we introduce two independent sets of the Gaussian basis functions $|(n_{\rho}l_{\rho}m_{\rho})^{G} \rangle$ and $|(n_{\lambda}l_{\lambda}m_{\lambda})^{G} \rangle$ based on the JC-3. Given a definite quantum state $|\{[(l_{\rho} l_{\lambda} )_{L}(s_{1}s_{2})_{s_{12}}]_{j} s_{3}\}_{JM_{J} }\rangle\equiv|\alpha\rangle_{3}$ (corresponding to the JC-3), the generalized Gaussian basis function has the form below,
\begin{eqnarray}
\notag
|(\tilde{n},\alpha)^{G}_{3}\rangle &&=\sum_{\{m_{\xi}\}}\{CG_{\xi}\}\times|(n_{\rho}l_{\rho}m_{\rho})^{G} \rangle\otimes|(n_{\lambda}l_{\lambda}m_{\lambda})^{G} \rangle\\
&&\otimes|s_{1}m_{s_{1}}\rangle\otimes|s_{2}m_{s_{2}}\rangle\otimes|s_{3}m_{s_{3}}\rangle,
\end{eqnarray}
where $\{m_{\xi}\}$ denote all the 3rd components of the orbital angular momenta and spins, $\{CG_{\xi}\}$ are the products of all the C-G coefficients. $\tilde{n}$ is obtained by combining $n_{\rho}$ and $n_{\lambda}$, e.g., $\tilde{n}=(n_{\rho}-1)\times n_{max}+n_{\lambda}$ as $n_{\rho(\lambda)}=1,\cdot\cdot\cdot,n_{max}$.
Thus, a given quantum state $(l_{\rho},l_{\lambda})L(J^{P})_{j}$ can be expanded in the radial space of the generalized Gaussian basis functions.

In the calculation of Hamiltonian matrix elements of three-body systems, particularly, when the Jacobi coordinates transformations are employed, integrations over all of the radial and angular coordinates become laborious even with the Gaussian basis functions. This process can be simplified by introducing the infinitesimally-shifted Gaussian (ISG) basis functions by
\begin{eqnarray}
\notag
\phi_{nlm}^{G} &&=N_{nl}r^{l}e^{-\nu_{n}r^{2}}Y_{lm}(\mathbf{\hat{r}})\\
&&=N_{nl}\lim_{\varepsilon\rightarrow~0}\frac{1}{(\nu_{n}\varepsilon)^{l}}\sum _{\tilde{k}=1}^{\tilde{k}_{max}}C_{lm,\tilde{k}}e^{-\nu_{n}(\mathbf{r}-\varepsilon \mathbf{D}_{lm,\tilde{k}})^{2}},
\end{eqnarray}
where, $r^{l}Y_{lm}(\mathbf{\hat{r}})$ is replaced by a set of coefficients $C_{lm,\tilde{k}}$ and vectors $\mathbf{D}_{lm,\tilde{k}}$. In this way, the Jacobi coordinates transformation just needs to be completed in the exponent section.
With the help of the ISG basis functions, the matrix elements of all the Hamiltonian terms can been rigorously calculated.

 Then, the energy levels of both the ground states and the radial excited states can be obtained by solving the generalized matrix eigenvalue problem in the framework of the GEM,
\begin{eqnarray}
\begin{aligned}
\sum^{n^{2}_{max}}_{\tilde{n}'=1}(H_{\tilde{n}\tilde{n}'}-EN_{\tilde{n}\tilde{n}'})C_{\tilde{n}'}=0.
\end{aligned}
\end{eqnarray}
Here $H_{\tilde{n}\tilde{n}'}$ is the matrix element of the Hamiltonian, $n_{max}^{2}$ is the total dimension of the radial space, and $N_{\tilde{n}\tilde{n}'}$ comes from the non-orthogonality of the Gaussian basis functions.

\section{Results and discussions}\label{sec3}

\begin{figure*}[htbp]
\centering
\includegraphics[width=21.5cm,trim=1cm 5cm 1cm 1cm]{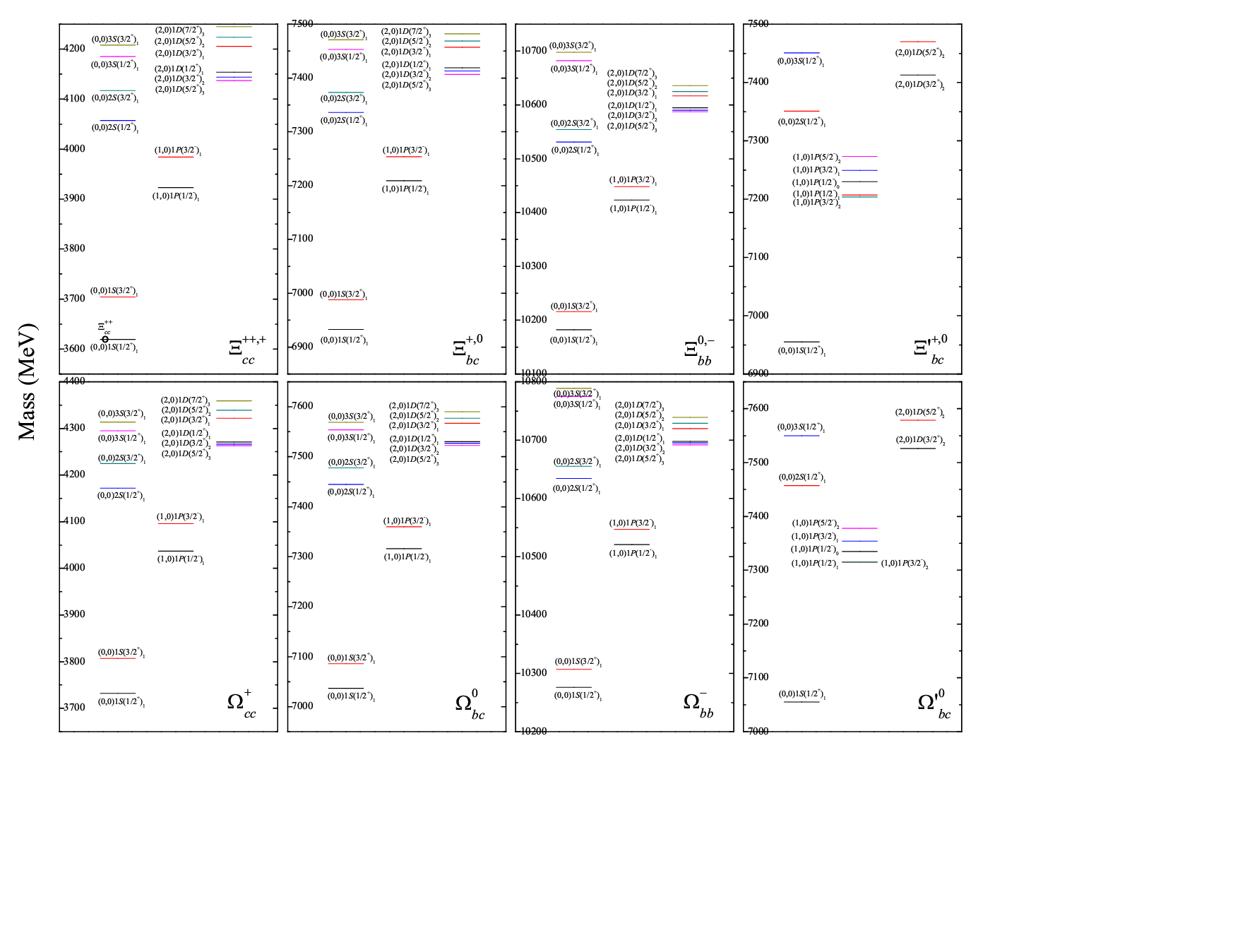}
\caption{Calculated spectra of the doubly heavy baryons and the relevant experimental data~\cite{04}. `++',`+', `0' and `-' indicate the charged states of baryons. The open circle denotes the $\Xi_{cc}^{++}$ baryon. }
\label{f2}
\end{figure*}

Based on the adopted methods mentioned above, the energy levels of the $1S$-, $2S$-, $3S$-, $1P$- and $1D$-wave states are obtained for all the DHB families, and their root-mean-square radii as well. Due to the fixed parameters $m_{u}=m_{d}=220$ MeV in this work, the calculated energy levels are degenerate for the isospin states, e.g., $m_{\Xi_{cc}^{++}}$ = $m_{\Xi_{cc}^{+}}$. The detailed calculation results are listed in Tables~\ref{a1}-~\ref{a3}, and the obtained spectral structure is presented in Fig.~\ref{f2}.

\begin{table*}[htbp]
\begin{ruledtabular}\caption{Contribution of each Hamiltonian term to the mass values (in MeV) for the $1S$-, $1P$- and $1D$-wave states of the $\Xi'_{bc}$, $\Xi_{bc}$ and $\Xi_{cc}$ baryons with $\langle H_{mode}\rangle\equiv\langle H_{0}+H^{conf}\rangle$ and $\langle H_{ij}\rangle\equiv\langle H\rangle-\langle(H-H_{ij})\rangle$. }
\label{a1}
\begin{tabular}{c c c c c c c c c c c c c c c c c c c}
$(l_{\rho},l_{\lambda})nL(J^{P})_{j}$ & $\langle H_{mode}\rangle$ & $\{\langle H^{t}_{12}\rangle$ & $\langle H^{t}_{23}\rangle$ & $\langle H^{t}_{31}\rangle\}$ & $\{\langle H^{c}_{12}\rangle$  & $\langle H^{c}_{23}\rangle$ & $\langle H^{c}_{31}\rangle\}$ & $\{\langle H^{SO(v)}_{12}\rangle$ & $\langle H^{SO(v)}_{23}\rangle$ & $\langle H^{SO(v)}_{31}\rangle\}$ & $\{\langle H^{SO(s)}_{12}\rangle$ & $\langle H^{SO(s)}_{23}\rangle$& $\langle H^{SO(s)}_{31}\rangle\}$& $\langle H\rangle$  \\\hline
\multicolumn{15}{c}{$\Xi'_{bc}$}   \\ \hline
$(0,0)1S(\frac{1}{2}^{+})_{0}$ & 6967.00  &$\{$ 0 & 0 & 0 $\}$&$\{$ -11.54  & 0 & 0 $\}$&$\{$ 0 & 0 & 0 $\}$&$\{$ 0 & 0 & 0 $\}$& 6955.47 \\
$(1,0)1P(\frac{1}{2}^{-})_{0}$ & 7240.25  & $\{$ 0 & 0 & 0 $\}$&$\{$ 0.37  & 0 & 0 $\}$&$\{$ -14.37 & -4.09 & -0.27 $\}$&$\{$ 4.76 & 3.29 & 0.16 $\}$& 7230.06  \\
$(1,0)1P(\frac{1}{2}^{-})_{1}$ & 7240.25  &$\{$ 0 & 0 & 0 $\}$&$\{$ 0.38  & -8.11 & -4.99 $\}$&$\{$ -7.38 & -15.37 & -2.37 $\}$&$\{$ 2.38 & 1.66 & 0.08 $\}$& 7207.19  \\
$(1,0)1P(\frac{3}{2}^{-})_{1}$ & 7240.25  &$\{$ -0.33 & 0.29 & 0.02 $\}$&$\{$ 0.34  & 3.82 & 2.34 $\}$&$\{$ -6.90 & 4.43 & 0.94 $\}$&$\{$ 2.35 & 1.62 & 0.08 $\}$& 7249.19  \\
$(1,0)1P(\frac{3}{2}^{-})_{2}$ & 7240.25  &$\{$ 0.47 & -0.19 & -0.01 $\}$&$\{$ 0.35  & -11.92 & -7.48 $\}$&$\{$ 7.25 & -17.62 & -3.20$\}$&$\{$ -2.36 & -1.65 & -0.08$\}$& 7204.02  \\
$(1,0)1P(\frac{5}{2}^{-})_{2}$ & 7240.25  &$\{$ 0.60 & 0.60 & 0.04 $\}$&$\{$ 0.30  & 7.20 & 4.50 $\}$&$\{$ 6.47 & 14.30 & 2.21$\}$&$\{$ -2.29 & -1.59 & -0.08$\}$& 7273.20  \\
$(2,0)1D(\frac{3}{2}^{+})_{2}$ & 7447.92  &$\{$ 0 & 0 & 0 $\}$&$\{$ -0.32  & 0 & 0 $\}$&$\{$ 0 & -29.00 & -5.93$\}$&$\{$ 0 & 0 & 0 $\}$& 7412.78  \\
$(2,0)1D(\frac{5}{2}^{+})_{2}$ & 7447.92  &$\{$ 0 & 0 & 0 $\}$&$\{$ -0.28  & 0 & 0 $\}$&$\{$ 0 & 18.72 & 3.83$\}$&$\{$ 0 & 0 & 0 $\}$& 7470.22  \\  \hline
\multicolumn{15}{c}{$\Xi_{bc}$}   \\ \hline
$(0,0)1S(\frac{1}{2}^{+})_{1}$ & 6967.00  &$\{$ 0 & 0 & 0 $\}$&$\{$ 3.79  & -26.15 & -12.50 $\}$&$\{$ 0 & 0 & 0 $\}$&$\{$ 0 & 0 & 0 $\}$& 6932.61  \\
$(0,0)1S(\frac{3}{2}^{+})_{1}$ & 6967.00  &$\{$ 0 & 0 & 0 $\}$&$\{$ 3.48  & 12.04 & 5.57 $\}$&$\{$ 0 & 0 & 0 $\}$&$\{$ 0 & 0 & 0 $\}$& 6988.35  \\
$(1,0)1P(\frac{1}{2}^{-})_{1}$ & 7240.25  &$\{$ 0 & 0 & 0 $\}$&$\{$ -1.07  & 0 & 0 $\}$&$\{$ 0 & -26.09 & -4.39 $\}$&$\{$ 0 & 0 & 0 $\}$& 7208.82  \\
$(1,0)1P(\frac{3}{2}^{-})_{1}$ & 7240.25  &$\{$ 0 & 0 & 0 $\}$&$\{$ -0.99  & 0 & 0 $\}$&$\{$ 0 & 12.78 & 2.14 $\}$&$\{$ 0 & 0 & 0 $\}$& 7254.18  \\
$(2,0)1D(\frac{1}{2}^{+})_{1}$ & 7447.92  &$\{$ 0 & 0 & 0 $\}$&$\{$ 0.11  & 5.24 & 4.13 $\}$&$\{$ -8.19 & -33.30 & -6.24 $\}$&$\{$ 5.06 & 4.35 & 0.22 $\}$& 7419.14  \\
$(2,0)1D(\frac{3}{2}^{+})_{1}$ & 7447.92  &$\{$ 0.04 & -0.45 & -0.04 $\}$&$\{$ 0.10  & -2.45 & -2.00 $\}$&$\{$ -7.70 & 10.15 & 2.58 $\}$&$\{$ 4.99 & 4.28 & 0.22 $\}$& 7457.47  \\
$(2,0)1D(\frac{3}{2}^{+})_{2}$ & 7447.92  &$\{$ -0.27 & -0.09 & 0 $\}$&$\{$ 0.11  & -2.60 & -2.08 $\}$&$\{$ -2.75 & -25.78 & -5.09 $\}$&$\{$ 1.68 & 1.45 & 0.08 $\}$& 7412.97  \\
$(2,0)1D(\frac{5}{2}^{+})_{2}$ & 7447.92  &$\{$ -0.35 & 0.30 & 0.03 $\}$&$\{$ 0.09  & 1.59 & 1.30 $\}$&$\{$ -2.49 & 14.27 & 3.08 $\}$&$\{$ 1.65 & 1.41 & 0.07 $\}$& 7468.89  \\
$(2,0)1D(\frac{5}{2}^{+})_{3}$ & 7447.92  &$\{$ 0.32 & -0.45 & -0.04 $\}$&$\{$ 0.10  & -6.83 & -5.55 $\}$&$\{$ 5.48 & -22.99 & -5.08 $\}$&$\{$ -3.35 & -2.90 &-0.15 $\}$& 7406.72  \\
$(2,0)1D(\frac{7}{2}^{+})_{3}$ & 7447.92  &$\{$ 0.33 & 0.87 & 0.08 $\}$&$\{$ 0.09  & 4.55 & 3.79 $\}$&$\{$ 4.78 & 20.93 & 3.97 $\}$&$\{$ -3.26 & -2.78 &-0.15 $\}$& 7481.78  \\\hline
\multicolumn{15}{c}{$\Xi_{cc}$}   \\ \hline
$(0,0)1S(\frac{1}{2}^{+})_{1}$ & 3670.69  &$\{$ 0 & 0 & 0 $\}$&$\{$ 7.38  & -30.26 & -30.26 $\}$&$\{$ 0 & 0 & 0 $\}$&$\{$ 0 & 0 & 0 $\}$ & 3618.76  \\
$(0,0)1S(\frac{3}{2}^{+})_{1}$ & 3670.69  &$\{$ 0 & 0 & 0 $\}$&$\{$ 6.58  & 13.18 & 13.18 $\}$&$\{$ 0 & 0 & 0 $\}$&$\{$ 0 & 0 & 0 $\}$ & 3704.28  \\
$(1,0)1P(\frac{1}{2}^{-})_{1}$ & 3965.92  &$\{$ 0 & 0 & 0 $\}$&$\{$ -2.15  & 0 & 0 $\}$&$\{$ 0 & -20.41 & -20.41 $\}$&$\{$ 0 & 0 & 0 $\}$& 3923.34  \\
$(1,0)1P(\frac{3}{2}^{-})_{1}$ & 3965.92  &$\{$ 0 & 0 & 0 $\}$&$\{$ -1.98  & 0 & 0 $\}$&$\{$ 0 & 9.89 & 9.89 $\}$&$\{$ 0 & 0 & 0 $\}$& 3983.74  \\
$(2,0)1D(\frac{1}{2}^{+})_{1}$ & 4195.15  &$\{$ 0 & 0 & 0 $\}$&$\{$ 0.20  & 7.40 & 7.40 $\}$&$\{$ -13.64 & -28.43 & -28.43 $\}$&$\{$ 8.28 & 3.21 & 3.21 $\}$& 4154.31  \\
$(2,0)1D(\frac{3}{2}^{+})_{1}$ & 4195.15  &$\{$ 0.07 & -0.33 & -0.33 $\}$&$\{$ 0.18  & -3.50 & -3.50 $\}$&$\{$ -13.00 & 8.48 & 8.48 $\}$&$\{$ 8.19 & 3.16 & 3.16 $\}$& 4205.99  \\
$(2,0)1D(\frac{3}{2}^{+})_{2}$ & 4195.15  &$\{$ -0.53 & -0.07 & -0.07 $\}$&$\{$ 0.19  & -3.71 & -3.71 $\}$&$\{$ -4.61 & -22.05 & -22.05 $\}$&$\{$ 2.76 & 1.07 & 1.07 $\}$& 4144.44  \\
$(2,0)1D(\frac{5}{2}^{+})_{2}$ & 4195.15  &$\{$ -0.69 & 0.21 & 0.21 $\}$&$\{$ 0.17  & 2.23 & 2.23 $\}$&$\{$ -4.17 & 11.98 & 11.98 $\}$&$\{$ 2.70 & 1.04 & 1.04 $\}$& 4224.21  \\
$(2,0)1D(\frac{5}{2}^{+})_{3}$ & 4195.15  &$\{$ 0.63 & -0.34 & -0.34 $\}$&$\{$ 0.19  & -9.79 & -9.79 $\}$&$\{$ 9.19 & -19.54 & -19.54 $\}$&$\{$ -5.48 & -2.14 &-2.14 $\}$& 4136.70  \\
$(2,0)1D(\frac{7}{2}^{+})_{3}$ & 4195.15  &$\{$ 0.65 & 0.62 & 0.62 $\}$&$\{$ 0.15  & 6.35 & 6.35 $\}$&$\{$ 7.93 & 17.54 & 17.54 $\}$&$\{$ -5.31 & -2.04 &-2.04 $\}$& 4244.99  \\
\end{tabular}
\end{ruledtabular}
\end{table*}

\begin{table*}[htbp]
\begin{ruledtabular}\caption{Calculated $\langle r_{\rho}^{2}\rangle^{1/2}$, $\langle r_{\lambda}^{2}\rangle^{1/2}$ (in fm) and mass values (in MeV) for the $1S$-, $2S$-, $3S$-, $1P$- and $1D$-wave states of the $\Xi(\Omega)_{cc}$, $\Xi(\Omega)_{bc}$ and $\Xi(\Omega)_{bb}$ baryons.}
\label{a2}
\begin{tabular}{c c c c c c c c c c c}
$(l_{\rho},l_{\lambda})nL(J^{P})_{j}$ & $\langle r_{\rho}^{2}\rangle^{1/2}$ & $\langle r_{\lambda}^{2}\rangle^{1/2}$ & $M_{cal.}$ & $\langle r_{\rho}^{2}\rangle^{1/2}$ & $\langle r_{\lambda}^{2}\rangle^{1/2}$ & $M_{cal.}$ & $\langle r_{\rho}^{2}\rangle^{1/2}$ & $\langle r_{\lambda}^{2}\rangle^{1/2}$ & $M_{cal.}$  \\ \hline
& \multicolumn{3}{c}{$\Xi_{cc}$}   &\multicolumn{3}{c}{$\Xi_{bc}$} &\multicolumn{3}{c}{$\Xi_{bb}$} \\\cline{2-4} \cline{5-7}  \cline{8-10}
$(0,0)1S(\frac{1}{2}^{+})_{0}$ & 0.424  & 0.454 & 3619  & 0.371  & 0.460 & 6933 & 0.295  & 0.464 & 10182\\
$(0,0)1S(\frac{3}{2}^{+})_{0}$ & 0.448  & 0.501 & 3704  & 0.385  & 0.492 & 6988 & 0.300  & 0.485 & 10216\\
$(0,0)2S(\frac{1}{2}^{+})_{0}$ & 0.741  & 0.595 & 4057  & 0.679  & 0.562 & 7336 & 0.580  & 0.522 & 10531\\
$(0,0)2S(\frac{3}{2}^{+})_{0}$ & 0.774  & 0.618 & 4117  & 0.700  & 0.571 & 7373 & 0.591  & 0.535 & 10554\\
$(0,0)3S(\frac{1}{2}^{+})_{0}$ & 0.532  & 0.833 & 4185  & 0.436  & 0.825 & 7453 & 0.313  & 0.837 & 10682\\
$(0,0)3S(\frac{3}{2}^{+})_{0}$ & 0.534  & 0.852 & 4208  & 0.428  & 0.844 & 7471 & 0.312  & 0.849 & 10698\\
$(1,0)1P(\frac{1}{2}^{-})_{1}$ & 0.638  & 0.512 & 3923  & 0.558  & 0.508 & 7209 & 0.455  & 0.498 & 10423\\
$(1,0)1P(\frac{3}{2}^{-})_{1}$ & 0.656  & 0.532 & 3984  & 0.575  & 0.521 & 7254 & 0.459  & 0.507 & 10448\\
$(2,0)1D(\frac{1}{2}^{+})_{1}$ & 0.814  & 0.551 & 4154  & 0.717  & 0.541 & 7419 & 0.583  & 0.523 & 10595\\
$(2,0)1D(\frac{3}{2}^{+})_{1}$ & 0.830  & 0.561 & 4206  & 0.736  & 0.546 & 7457 & 0.588  & 0.529 & 10617\\
$(2,0)1D(\frac{3}{2}^{+})_{2}$ & 0.813  & 0.545 & 4144  & 0.717  & 0.537 & 7413 & 0.582  & 0.519 & 10590\\
$(2,0)1D(\frac{5}{2}^{+})_{2}$ & 0.846  & 0.573 & 4224  & 0.748  & 0.553 & 7469 & 0.593  & 0.534 & 10625\\
$(2,0)1D(\frac{5}{2}^{+})_{3}$ & 0.819  & 0.543 & 4137  & 0.721  & 0.535 & 7407 & 0.584  & 0.516 & 10587\\
$(2,0)1D(\frac{7}{2}^{+})_{3}$ & 0.866  & 0.586 & 4245  & 0.763  & 0.561 & 7482 & 0.599  & 0.539 & 10636\\\hline
& \multicolumn{3}{c}{$\Omega_{cc}$}   &\multicolumn{3}{c}{$\Omega_{bc}$} &\multicolumn{3}{c}{$\Omega_{bb}$} \\\cline{2-4} \cline{5-7}  \cline{8-10}
$(0,0)1S(\frac{1}{2}^{+})_{0}$ & 0.417  & 0.421 & 3732  & 0.365  & 0.422 & 7037 & 0.290  & 0.422 & 10276\\
$(0,0)1S(\frac{3}{2}^{+})_{0}$ & 0.439  & 0.461 & 3807  & 0.379  & 0.450 & 7086 & 0.296  & 0.440 & 10307\\
$(0,0)2S(\frac{1}{2}^{+})_{0}$ & 0.715  & 0.575 & 4172  & 0.662  & 0.533 & 7445 & 0.568  & 0.483 & 10634\\
$(0,0)2S(\frac{3}{2}^{+})_{0}$ & 0.744  & 0.595 & 4225  & 0.684  & 0.538 & 7478 & 0.579  & 0.493 & 10655\\
$(0,0)3S(\frac{1}{2}^{+})_{0}$ & 0.547  & 0.774 & 4295  & 0.447  & 0.765 & 7554 & 0.315  & 0.780 & 10775\\
$(0,0)3S(\frac{3}{2}^{+})_{0}$ & 0.549  & 0.791 & 4314  & 0.436  & 0.785 & 7569 & 0.313  & 0.791 & 10789\\
$(1,0)1P(\frac{1}{2}^{-})_{1}$ & 0.625  & 0.473 & 4037  & 0.548  & 0.466 & 7316 & 0.447  & 0.454 & 10521\\
$(1,0)1P(\frac{3}{2}^{-})_{1}$ & 0.645  & 0.495 & 4096  & 0.566  & 0.481 & 7360 & 0.452  & 0.465 & 10547\\
$(2,0)1D(\frac{1}{2}^{+})_{1}$ & 0.798  & 0.509 & 4271  & 0.704  & 0.497 & 7530 & 0.573  & 0.479 & 10698\\
$(2,0)1D(\frac{3}{2}^{+})_{1}$ & 0.818  & 0.525 & 4322  & 0.725  & 0.505 & 7567 & 0.579  & 0.487 & 10720\\
$(2,0)1D(\frac{3}{2}^{+})_{2}$ & 0.800  & 0.507 & 4266  & 0.707  & 0.494 & 7527 & 0.573  & 0.476 & 10695\\
$(2,0)1D(\frac{5}{2}^{+})_{2}$ & 0.834  & 0.536 & 4339  & 0.737  & 0.512 & 7577 & 0.584  & 0.491 & 10729\\
$(2,0)1D(\frac{5}{2}^{+})_{3}$ & 0.809  & 0.507 & 4263  & 0.712  & 0.493 & 7523 & 0.575  & 0.474 & 10692\\
$(2,0)1D(\frac{7}{2}^{+})_{3}$ & 0.855  & 0.548 & 4359  & 0.752  & 0.519 & 7590 & 0.591  & 0.496 & 10739\\
\end{tabular}
\end{ruledtabular}
\end{table*}

\begin{table*}[htbp]
\begin{ruledtabular}\caption{Same as Table~\ref{a2}, but for the $\Xi'_{bc}$ and $\Omega'_{bc}$ baryons.}
\label{a3}
\begin{tabular}{c c c c c c c c c c c}
$(l_{\rho},l_{\lambda})nL(J^{P})_{j}$ & $\langle r_{\rho}^{2}\rangle^{1/2}$ & $\langle r_{\lambda}^{2}\rangle^{1/2}$ & $M_{cal.}$ & $\langle r_{\rho}^{2}\rangle^{1/2}$ & $\langle r_{\lambda}^{2}\rangle^{1/2}$ & $M_{cal.}$ \\ \hline
& \multicolumn{3}{c}{$\Xi'_{bc}$}   &\multicolumn{3}{c}{$\Omega'_{bc}$} \\ \cline{2-4} \cline{5-7}
$(0,0)1S(\frac{1}{2}^{+})_{1}$ & 0.370  & 0.479 & 6955  & 0.363  & 0.438 & 7055  \\
$(0,0)2S(\frac{1}{2}^{+})_{1}$ & 0.683  & 0.569 & 7351  & 0.666  & 0.538 & 7457  \\
$(0,0)3S(\frac{1}{2}^{+})_{1}$ & 0.422  & 0.834 & 7451  & 0.433  & 0.774 & 7550   \\
$(1,0)1P(\frac{1}{2}^{-})_{0}$ & 0.557  & 0.513 & 7230  & 0.547  & 0.472 & 7335  \\
$(1,0)1P(\frac{1}{2}^{-})_{1}$ & 0.552  & 0.503 & 7207 & 0.543  & 0.463 & 7315  \\
$(1,0)1P(\frac{3}{2}^{-})_{1}$ & 0.569  & 0.521 & 7249  & 0.560  & 0.480 & 7354 \\
$(1,0)1P(\frac{3}{2}^{-})_{2}$ & 0.560  & 0.502 & 7204  & 0.552  & 0.462 & 7315 \\
$(1,0)1P(\frac{5}{2}^{-})_{2}$ & 0.589  & 0.531 & 7273  & 0.580  & 0.489 & 7378 \\
$(2,0)1D(\frac{3}{2}^{+})_{2}$ & 0.720  & 0.539 & 7413  & 0.496  & 0.928 & 7526  \\
$(2,0)1D(\frac{5}{2}^{+})_{2}$ & 0.751  & 0.553 & 7470  & 0.740  & 0.512 & 7579  \\
\end{tabular}
\end{ruledtabular}
\end{table*}

(1) Orbital excitation mode in the HQD mechanism

For correctly describing the structures of the excitation spectra, the HQD mechanism of the orbital excitation was proposed in Ref.~\cite{96}, which has been testified to be reasonable for singly and doubly heavy baryons. Recently, this mechanism was successfully applied to the study of the singly heavy baryons and solved the problem of the `missing' states in a natural way. In this work, the HQD mechanism is applied to the DHBs. The results show that the orbital excitation of the DHBs is dominated by the $\rho$-mode ($l_{\lambda}=0$, see Tables~\ref{a1}-~\ref{a3}), i.e., the dominant orbital excitation only occurs between the two heavy quarks as discussed in Refs.~\cite{F403,96}.

In some earlier works with the heavy-diquark approximation, the orbital excitation of the DHBs was assumed to occur between the heavy-diquark and the light quark, where the heavy-diquark was degraded to a point-like particle~\cite{104,105,106}. Later, Roberts pointed out that the predicted excited energies by these works are too large, while the excited energies from the orbital excitation only contributed by the two heavy quarks are relatively low~\cite{F403}. On the other hand, Ebert et al. took both of these two excitations into account~\cite{62p3}, and the number of the predicted excited states increases significantly.

In this work, the orbital excitation only from the two heavy quarks is taken into account. As a result, not only are the excitation energies relatively low, but also the predicted number of the excited states is small. Ultimately, the correctness of the predicted spectral structure still needs to be verified by the future experiments.

(2) Contribution of each Hamiltonian term to the energy levels.

The contribution of each Hamiltonian term to the energy levels of the $\Xi'_{bc}$, $\Xi_{bc}$ and $\Xi_{cc}$ families is presented in Table~\ref{a1}.
Among the expectation values of these Hamiltonian terms, $\langle H_{mode}\rangle\equiv\langle H_{0}+H^{conf}\rangle$ depends on the excitation modes ($l_{\rho}, l_{\lambda}$) and dominates the main part of the energy levels. The others affect the shifts and splittings of the energy levels.

As is shown in Table~\ref{a1}, the expectation values of the tensor terms $\langle H^{t}_{ij}\rangle$ have little influence on the energy levels. The expectation values of interactions between the two heavy quarks ($\langle H^{c}_{12}\rangle$, $\langle H^{SO(v)}_{12}\rangle$ and $\langle H^{SO(s)}_{12}\rangle$) shift the energy levels by several MeV, nevertheless, have little effect on the energy level splittings. Especially, for the $\Xi'_{bc}$ family, $\langle H^{c}_{12}\rangle$ alone causes the $S$-wave state a certain energy level.

For the orbital excited states, the influence on the energy level splitting of the spin doublet states with total angular momenta ($j+\frac{1}{2}$, $j-\frac{1}{2}$) mainly comes from $\langle H^{SO(v)}_{23}\rangle$ and $\langle H^{SO(v)}_{31}\rangle$. For the spin doublet states ($\frac{1}{2}$, $\frac{3}{2}$) of the $S$-wave in the $\Xi_{bc}$ and $\Xi_{cc}$ families, however, their energy level splittings are significantly determined by the expectation values of the contact terms $\langle H^{c}_{23}\rangle$ and $\langle H^{c}_{31}\rangle$.

In short, the relativistic motion mass $H_{0}$ and the position-related interactions $H^{conf}$ determine the main part of the energy levels. The spin-dependent interactions between the two heavy quarks ($H^{hyp}_{12}$ and $H^{SO}_{12}$) only shift the energy levels, while the energy level splittings mainly result from the ones between the heavy quarks and the light quark ($H^{hyp}_{23(31)}$ and $H^{SO}_{23(31)}$). In other words, the spin-splittings are mainly determined by the spin-dependent interactions associated with the light quark.

(3) Structures of the spectra.

For the $\Xi_{cc}$, $\Xi_{bc}$ and $\Xi_{bb}$ families, their energy level structures are the same, and the energy splitting of their corresponding spin doublet states ($j-\frac{1}{2}$,$j+\frac{1}{2}$) evolves gradually with the heavy quark mass. As is shown in Fig.~\ref{f2}, for example, the energy gap between the $1S(\frac{1}{2}^{+})_{1}$ and $1S(\frac{3}{2}^{+})_{1}$ states becomes smaller slowly as the mass of the heavy quark increases. This feature conforms to the prediction of the heavy quark limit and heavy quark symmetry~\cite{06}. The same feature exists in the $\Omega_{cc}$, $\Omega_{bc}$ and $\Omega_{bb}$ families, but their energy gap is generally smaller. This is caused by the different masses of the light quarks.
In the $\Xi'_{bc}$ and $\Omega'_{bc}$ families, their spectra present a different structure instead, due to the antisymmetry of their flavor wave functions.

There is also an interesting property hidden in Fig.~\ref{f2}, i.e., the Regge trajectories. One can see that the energy levels of some related states almost stand in a line, such as the $(0,0)1S(\frac{1}{2}^{+})_{1}$, $(1,0)1P(\frac{3}{2}^{-})_{1}$ and $(2,0)1D(\frac{5}{2}^{+})_{2}$ states in the $\Xi_{cc}$, $\Xi_{bc}$ and $\Xi_{bb}$ families, or the $(0,0)1S(\frac{1}{2}^{+})_{1}$, $(1,0)1P(\frac{3}{2}^{-})_{2}$ and $(2,0)1D(\frac{5}{2}^{+})_{2}$ states in the $\Xi'_{bc}$ family, which are grouped with the natural parity $P=(-1)^{J-\frac{1}{2}}$~\cite{F405}.

(4) Searching for the DHBs.

As is discussed in Sec.~\ref{sec1}, the study of the DHBs requires reliable theoretical calculations. The combination method adopted in this work has been proved to be effective and reliable in the study of the singly heavy baryons, where the average accuracy is estimated to be less than 10 MeV~\cite{101}. In this work, for the $1S(\frac{1}{2}^{+})$ state of the double charm baryons, the predicted mass is 3619 MeV, which is very close to the central value of the data 3621.6 MeV~\cite{04} and the results of the lattice QCD~\cite{44,46p}. From this, it can be estimated that the uncertainty in this work should also be less than 10 MeV.

As is shown in Table~\ref{a01}, the ground state masses given by many theoretical works including this work are more than 3600 MeV. This suggests that the $\Xi_{cc}^{+}(3520)$ state should not exist truly. And the $\Xi_{cc}^{++}(3621)$ should be the true ground state with $J^{P}$ = $\frac{1}{2}^{+}$. For its spin-doublet partner $1S(\frac{3}{2}^{+})$ state, the calculated mass value of the double charm baryons is 3704 MeV as shown in Table~\ref{a2}, which is consistent with the prediction of the lattice QCD in Table~\ref{a01}. On the other hand, the experimental research shows that there may be an indistinct small peak near 3695 MeV (see the Fig.1 in Ref.~\cite{14} and Fig.2 in Ref.~\cite{15}). Thus, it is suggested that the relevant experiment could be carefully designed to search for the $\Xi_{cc}^{*}$ in the energy range from 3694 to 3714 MeV.

Before the $\Xi_{cc}^{++}(3621)$ was discovered by the LHCb collaboration, Ref.~\cite{27} had suggested that the decay mode $\Xi_{cc}^{+}\rightarrow\Xi_{c}^{++}(2520)K^{-}$ could be reconstructed in the $\Lambda_{c}^{+}K^{-}\pi^{+}$ final states which might be easy to access. Now, we believe that the favorable decay modes of the $\Xi_{cc}^{*++}$ are still the $\Lambda_{c}^{+}K^{-}\pi^{+}\pi^{+}$ and $\Xi_{c}^{+}\pi^{+}$~\cite{14,15}, which should take precedence over the search for the $\Xi_{cc}^{+}$ and $\Xi_{cc}^{*+}$ with the final states $\Lambda_{c}^{+}K^{-}\pi^{+}$ and $pD^{+}K^{-}$, as discussed in Ref.~\cite{28}. Recently, it is pointed out that the process $\Xi_{cc}^{+}\rightarrow\Lambda_{c}^{+}\pi^{+}\pi^{0}K^{-}$ would be helpful in searching for the $\Xi_{cc}^{+}$, and $\Omega_{cc}^{+}\rightarrow\Lambda_{c}^{+}\pi^{+}\bar{K}^{0}K^{-}$ for the $\Omega_{cc}^{+}$~\cite{34p}.

The predicted bottom-charm baryon masses of the $1S$-wave states in this work are $M_{\Xi_{bc}}$ = 6933 MeV, $M_{\Xi'_{bc}}$ = 6955 MeV and $M_{\Xi_{bc}^{*}}$ = 6988 MeV, respectively. With the lattice QCD~\cite{44}, they are 6943 MeV, 6959 MeV and 6985 MeV, respectively. The results calculated by the two methods are also very consistent with each other. In~\cite{27}, the analysis showed that the sizeable branching ratio of $\Xi_{bc}^{0}\rightarrow pK^{-}$ is expected to be observed in experiments. However, the LHCb collaboration reported that no significant signal is found in the invariant mass range from 6.7 to 7.2 GeV$/c^{2}$ in searching for the bottom-charm baryons~\cite{16,18}. In the search of the $\Omega_{bc}^{0}$, the signal of interest was also not observed for the invariant masses between 6700 and 7300 MeV$/c^{2}$~\cite{17}. While, the predicted masses in this work ($M_{\Omega_{bc}}$ = 7037 MeV, $M_{\Omega'_{bc}}$ = 7055 MeV and $M_{\Omega_{bc}^{*}}$ = 7086 MeV) are exactly located within this mass range. One of the possible reasons for these zero results in the experiments might be that the statistics of the events are rather poor. The second reason for this may be the shorter life-times that are expected for these states. The 3rd one may lie in the small production rate.
In~\cite{108p}, the production rate of the $\Xi_{cc}$, the cross section of the $\Xi_{bc}$ at LHCb and the lifetimes of the DHBs were analyzed in detail. The result showed that the $\Xi_{bc}$ might be observed in the LHCb data of Run III. As for the observation of the $\Xi_{bb}$, however, the authors doubted its possibility at the LHCb because of the very small production rate.

The above difficulties in experimental observations are expected to be overcome in the updated colliders, such as the high luminosity LHC (HL-LHC), by which the improved vertex resolution of the upgraded LHCb detector together with larger data samples, will provide a strong support for the discovery of the DHBs~\cite{01}.
Very recently, the production of the DHBs was studied in the heavy ion ultra-peripheral collisions (UPCs)~\cite{108}. And the numerical results indicated that the experimental investigation for the $\Xi_{cc}$ baryons is feasible at the coming HL-LHC.

In addition, there are many theoretical studies in predicting the production rates, the weak decay modes, the strong decay modes and the lifetimes of the DHBs, as mentioned in Sec.~\ref{sec1}. These references and the references within them can serve as theoretical supports for the related experiments.

\section{Conclusions}\label{sec4}

In this work, the combination method of the RQM + HQD + GEM(ISG) is used to rigorously calculate the DHBs mass spectra in the three-quark system, which has been proved to be successful and reliable in the study of the singly heavy baryons.
Under the HQD mechanism, the calculated result shows that the orbital excitation of the DHBs is dominated by the $\rho$-mode. The excitation spectra of all DHBs families are systematically obtained and analyzed.

The energy level splitting of the spin-doublet states is found to evolve regularly along with the heavy-quark masses, which agrees with the conclusion of the heavy-quark symmetry.
By analyzing the contribution of each Hamiltonian term to the energy levels, we find that the spin-splitting is mainly determined by the spin-dependent interactions associated with the light quark ($H^{hyp}_{23(31)}$ and $H^{SO}_{23(31)}$).

The calculation in this work, along with some other theoretical studies, suggests that the $\Xi_{cc}^{+}(3520)$ state should not exist truly and the $\Xi_{cc}^{++}(3621)$ state should be the true ground state with $J^{P}$ = $\frac{1}{2}^{+}$.  By comparing the experimental data from the LHCb collaboration with the theoretical results given by this work and the lattice QCD, we suggest that the relevant experiment could be carefully designed to search for the $\Xi_{cc}^{*}$ state in the energy range from 3694 to 3714 MeV.

Based on the study of the singly heavy baryons with the same combination method, the calculation uncertainty in this work is estimated to be less than 10 MeV. The obtained DHBs mass spectra in this work may provide valuable references for related researches.

\begin{large}
\section*{Acknowledgements}
\end{large}

This research was supported by the Open Project of Guangxi Key Lab of Nuclear Physics and Technology (No. NLK2023-04), the Central Government Guidance Funds for Local Scientific and Technological Development in China (No. Guike ZY22096024), the Natural Science Foundation of Guizhou Province-ZK[2024](General Project)650, the National Natural Science Foundation of China (Grant Nos. 11675265, 12175068), the Continuous Basic Scientific Research Project (Grant No. WDJC-2019-13) and the Leading Innovation Project (Grant No. LC 192209000701).

\end{document}